# A review for compact model of graphene field-effect transistors

Nianduan Lu(卢年端) [1,2,3], Lingfei Wang(汪令飞) [1,3], Ling Li(李泠) [1,2,3] and Ming Liu(刘明)[1,2,3]

[1] *Key Laboratory of Microelectronic Devices & Integrated Technology, Institute of Microelectronics of Chinese Academy of Sciences, Beijing 100029, China*
[2] *University of Chinese Academy of Sciences, Beijing 100049, China*
[3] *Jiangsu National Synergetic Innovation Center for Advanced Materials (SICAM), Nanjing, 210009, China*

**Abstract**

Graphene has attracted enormous interests due to its unique physical, mechanical, and electrical properties. Specially, graphene-based field-effect transistors (FETs) have evolved rapidly and are now considered as an option for conventional silicon devices. As a critical step in the design cycle of modern IC products, compact model refers to the development of models for integrated semiconductor devices for use in circuit simulations. The purpose of this review is to provide a theoretical description of current compact model of graphene field-effect transistors. Special attention is devoted to the charge sheet model, drift-diffusion model, Boltzmann equation, density of states (DOS), and surface-potential-based compact model. Finally, an outlook of this field is briefly discussed.

## 1. Introduction

As a two-dimensional material with honeycomb structure, graphene has attracted enormous interests because of its unique properties, since its debut in 2004 [1-3]. Its unique physical, mechanical, and electrical properties have drawn a lot of interests among scientists [4-6]. Graphene not only exhibits excellent optoelectronic and mechanical properties but also can provide good adhesion with several organic materials so as to produce high-performance organic field-effect transistors [7-10]. As compared with conventional semiconductors, such as silicon, graphene shows completely different properties, for example, it is a zero-overlap semiconductor with very high electrical conductivity, and its conduction and valence bands meet at the Dirac point. For semiconductors, the flow of electricity requires some kinds of activation (such as heat or light absorption) to get over the gap between the valence band and conduction band. If the semiconductor is activated by the external electric field to switch "on" and "off", then it is called field-effect transistors (FETs) [11, 12]. A FET consists of a gate, a channel region connecting source and drain electrodes, and a barrier separating the gate from the channel, as shown in Fig. 1(a). The operation of a conventional FET relies on the control of the channel conductivity, and thus the drain current, by a voltage ($V_{GS}$) applied between the gate and source. The corresponding transfer characteristics curve of FET is shown in Fig. 1(c). Generally speaking, the large-scale and bilayer graphene do not possess a band gap. However, constraining large-scale graphene in one dimension or applying an electric field perpendicularly on the bilayer graphene can induce the band gap [13-15]. Recently, both theoretical and experimental works have achieved the graphene FETs [16-18]. Fig. 1(b) shows a structure of top-gated graphene FET (GFET). The corresponding transfer characteristics curve for two GFETs with large-area-graphene channel is shown in Fig. 1(d). One can see in Fig. 1 that, due to the different structure between conventional MOSFET and GFET, the transfer characteristics curve shows entirely different features. For example, the transfer characteristics curve of GFETs have two kinds of conductions, i.e., hole conduction and electron conduction, while MOSFET has a unique current-voltage transfer characteristic.

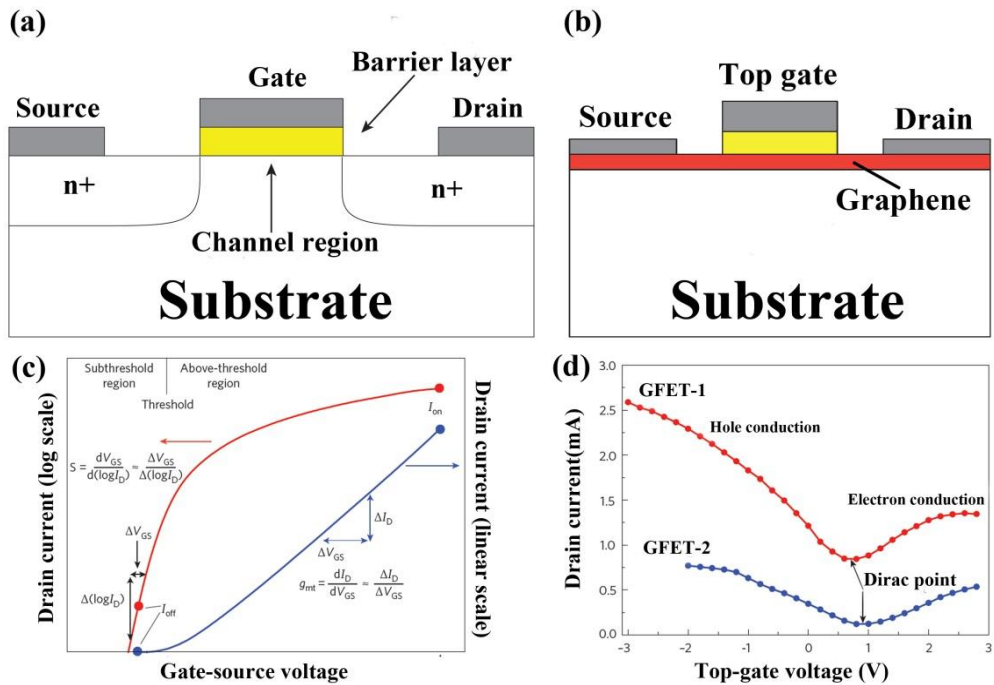

Fig. 1 Cross-section of a n-channel MOSFET (a) and a top-gated graphene FET (b), (c) MOSFET transfer characteristics showing $I_D$ (on a logarithmic scale on the left and a linear scale on the right) versus the gate-source voltage ($V_{GS}$), and (d) transfer characteristics curve for two GFETs with large-area-graphene channel.

Semiconductor device model, being a connection bridge between semiconductor manufactures and circuit designers, plays a critical role in semiconductor industry [19, 20]. With the development of CMOS technology which brings plenty of new physical effects, semiconductor device model has rapidly developed. Currently, integrated circuit (IC) designers use various kinds of software (such as SPICE, SLIC, PHILIPAC) in circuit design [21-23]. The core of the corresponding software is the model of each unit device. Because integrated circuit includes plenty of transistors, if all unit devices use the complicated model of transistor, the system level simulation will beyond computer ability and hence results in non-convergence in calculation. On the other hand, the transistor model can accurately describe its physical properties in order to ensure the reliability of the calculated result.

It is reasonable to state what a compact model is, since one may see differences in interpretations in the past. The following requirements for an excellent compact model will be considered in Ref. [24, 25].

(i) Must represent consistently the behavior of FET device.

(ii) Must be symmetrical to reflect the symmetry of FET structure.

(iii) Has to be analytical, without differentials or integrals. Could be in a matrix format but must not be a finite element model.

(iv) Has to be simple and easily derivable. No compromises with the overall FET behavior but some freedom for simplifications should be allowed, neglecting details that deviate from experimental trends, are insignificant in magnitude, or are not reproducible in identical devices and measurements. In other words, one should have a generic model.

(v) Has to have parameters that can be characterized relatively easy, or even guessed.

(vi) Has to be upgradable and reducible. Consequently, different portions and dependences (e.g., on temperature) in the model can be replaced with different models. Thus, the different portions in the model are extensions of each other, and the portions communicate between each other only by the values of the parameters or the quantities but the relations in the different portions must remain separated.

(vii) Has to have relations that can be physically justified.

(viii) Should have similar form and correspondence to compact models for other FETs.

(ix) Should be tunable to inaccurate (or uncertain) experimental data. Consequently, scaling rules can be left external, since these rules can be obscured by individual devices and different measurements.

Compact model is a critical step in the design cycle of modern IC products [26]. It refers to the development of models for integrated semiconductor devices for use in circuit simulations. The models are used to reproduce device terminal behaviors with accuracy, computational efficiency, ease of parameter extraction, and relative model simplicity for a circuit or system-level simulation, for future technology nodes [27]. Physics-based models are often preferred, particularly when concerned with statistical or predictive simulation. The industry's dependence on accurate and time-efficient compact models continues to grow as circuit operating frequencies increase and device tolerances scale down with concomitant increases in chip device count, and analog content in mixed-signal circuits. Accurate and physics-based compact models are useful for the design and development of FETs for digital and analog circuits. These models are highly desirable because they offer better computational efficiency than their numerical alternatives without loss of physical insights. Since T. Kacprzak et al. had reported firstly the compact DC model of GaAs-FETs in 1983 [28], the compact model for FETs has received much attention [29-33]. Fig. 2 shows the Thomson Reuters Web of Science publication report for the topic "Compact Model for Field-Effect Transistors" in the last 20 years and "Compact Model for Graphene FET". Research interest in the compact model has been growing remarkably over the last 10 years. Otherwise, the earliest compact model for graphene FET was reported by Meric et al. in 2008 [34, 35], which is based on the charge sheet model of MOSFET. Following Meric's model, several evolving compact models have been established [36-39], such as, virtual-source current-voltage model, SPICE-compatible compact model, and electrical compact model, etc.. Otherwise, with the decrease of channel lengths to below 50 nm, traditional compact models lose validity. To satisfy the scale-down requests, the quasi ballistic FET model has been developed for nano scale FETs [40]. However, due to the difference of structure and transport feature between MOSFET and graphene FETs (GFETs), the model of MOSFET maybe not entirely practicable to GFETs. Some new compact models, for example, based on a drift-diffusion model [41-45] and Boltzman equation [46], have also been developed. Further, some new physical-based compact models, such as surface-potential-based [47, 48] and based on density of states (DOS) [49-51], have been developed to achieve high accuracy and more physical.

**(a)**                                                    **(b)**

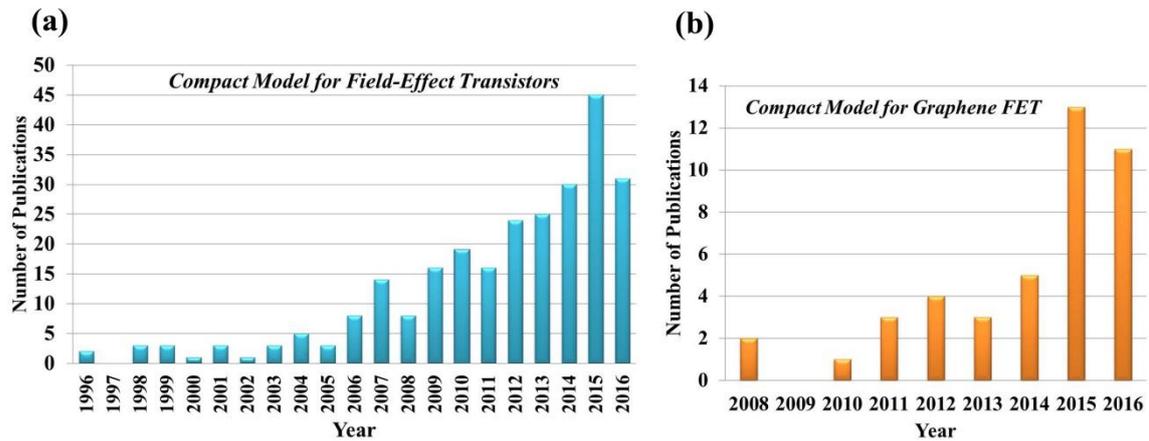

Fig. 2 Thomson Reuters Web of Science publication report for the topic, (a) "Compact Model for Field-Effect Transistors" in the last 20 years, and (b) "Compact Model for Graphene FET".

There have been some excellent reviews published recently with emphasis on the basic science of graphene and graphene FETs [7, 11, 52-57]. Given the growing interest in graphene in the electron-device community, and ongoing discussions of the potential of graphene transistors, a review article focusing on compact model of graphene FETs is timely. In this review, we mainly focus on the compact model of graphene FETs based on different methods, including not only based on conventional MOSFET model but also new physical-based model. In Section 2, the theoretical basis of compact model is discussed. In Section 3-7, the charge sheet model, drift-diffusion model, Boltzmann equation, surface-potential-based compact model and the model based on density of states (DOS) are summarized. Finally, the future outlook for this field is briefly discussed in Section 8.

## 2. Theoretical basis of compact model

As mentioned above, the first compact model for FETs was proposed by T. Kacprzak et al. in 1983 [28]. After more than 30 years, several researchers have developed various compact models in FETs. All these proposed compact models can be divided into two categories, one is charge-based and another is surface-potential-based. Next, we will describe the theoretical basis of compact model developed from charge-based and surface-potential-based, respectively.

## 2.1 Charge sheet model

The charge sheet model is applied to long-channel devices. As compared with other models, such as Pao-Sah model [58], the charge sheet model leads to a very simple algebraic formula for the current of long-channel devices, which can be used in all regimes from subthreshold to saturation. The charge sheet model firstly assumes a quasi-Fermi level formulation for carrier densities and the coordinate system. The source-to-drain current, $I$, can be related to the average quasi-Fermi level gradient, $d\overline{\Phi}_f/dy$, and the carrier density per unit area, $N(y)$, [59]

$$I = Zq\mu N(y)\, d\overline{\Phi}_f/dy, \tag{1}$$

where Z is the channel width, $\mu$ is the effective mobility, $q$ is the elementary charge, and $y$ is the measured distance along the channel from the source toward the drain, $N(y) = \int_0^\infty dx[n(x,y) - n_f(x,y)]$. $n - n_f$ is the minority carrier density per unit volume in excess of the zero band-banding density $n_f$. $d\overline{\Phi}_f/dy$ can be simply approximated by

$$\frac{d\overline{\Phi}_f}{dy} = \frac{d\Phi_s}{dy} - \frac{1}{\beta}\frac{d\ln N}{dy}, \tag{2}$$

where $\Phi_s(y)$ is the potential along oxide-silicon interface and $\beta = \frac{q}{kT}$. By integrating Eq. (2), one can obtain

$$N(y) = N(0)exp\{\beta[\Phi_s(y) - \Phi_s(0)] - \beta[\overline{\Phi}_f(y) - \overline{\Phi}_f(0)]\}. \tag{3}$$

That is, Eq. (2) is equivalent to the assumption that the carrier density along the channel varies only because the inversion layer moves rigidly with respect to the quasi-Fermi level as the potential varies. All shape dependence of the inversion layer upon carrier density or normal field is contained in $N(0)$ and remains unaltered from source to drain.

By substituting Eq. (2) into Eq. (1), one can solve Eq. (1) for $N(y)$ to obtain,

$$N(y) = N(0)\, exp\{\beta[\Phi_s(y) - \Phi_s(0)]\}$$
$$- \frac{I}{kT\mu Z}exp(\beta\Phi_s(y))\int_0^y dy_0 exp(-\beta\Phi_s(y_0)). \tag{4}$$

The model is then addressed by using Poisson's equation for the potential,

$$\begin{cases} \nabla^2 \Phi = 0 & (in\ gate\ oxide) \\ \nabla^2 \Phi = -\frac{q(p-N_A)}{\kappa \epsilon_0} & (in\ silicon) \end{cases}, \qquad (5)$$

where one can assume $p$-type material. Eq. (5) is joined across the oxide-silicon interface by the discontinuity condition

$$\kappa_{ox} \epsilon_0 \frac{\partial \Phi}{\partial x}\Big|_{o^-} - \kappa_{sc} \epsilon_0 \frac{\partial \Phi}{\partial x}\Big|_{o^+} = -qN(y), \qquad (6)$$

where $(o^-)$ refers to evaluation on the oxide side of the interface, $(o^+)$ on the silicon side. The $x$-coordinate measures the normal distance from the interface to the silicon. Eq. (6) implies that $N(y)$ is contained in a charge sheet of zero thickness. This in turn means (i) the current is constrained to flow along the oxide-silicon inter-face, and (ii) there is no voltage drop across the inversion layer. Using some arbitrary $\Phi_s(y) \equiv \Phi(x = 0, y)$, one can solve Eq. (5) for $\Phi$ in both oxide and semiconductor. This determines the normal derivatives in Eq. (6). Using Eq. (4), one can obtain an integral equation for $\Phi_s(y)$. This equation now will be set up and solved for long channel devices.

## 2.2 Boltzmann equation

In order to model the transport characteristics of the GFET, one can split carrier distribution function into its even and odd parts, that is, $f(\vec{k}) = f_{even}(\vec{k}) + f_{odd}(\vec{k})$. Then, it is well known that in the presence of randomizing collisions, and even in high fields, the Boltzmann transport equation can be written as[46, 60]

$$\frac{qF}{\hbar} \frac{\partial}{\partial k_x} f_{even}(\vec{k}) = -\frac{1}{\tau_{tot}(k)} f_{odd}(\vec{k}), \qquad (7)$$

with $\frac{1}{\tau_{tot}(k)} = \sum_i \frac{1}{\tau_i(k)}$ and the $i$-index indicates a particular scattering mechanism. In the presence of strong inter-carrier scattering for high carrier concentration, the even part of the distribution is thermalized at an electronic temperature $T_e$, and reads as

$$f_{even}(\vec{k}) = \frac{1}{1+exp[(\hbar v_F)(k-k_F)/(k_B T_e)]}, \qquad (8)$$

where $k_F = k_F(x)$ defines the carrier concentration along the channel. In $p$-channel, the current can be calculated as

$$\vec{I} = \frac{4q}{L} \sum_{\vec{k}} \vec{v}(\vec{k}) f_{odd}(\vec{k}), \qquad (9)$$

where $L$ is the channel length, and the factor 4 accounts for the spin and the twofold degeneracy of the Dirac point. Here, $v(\vec{k}) = vF(\cos\theta, \sin\theta)$ and $\theta$ is the angle between the electric field and the vector $\vec{k}$. Then for $\hbar v_F k_F \gg k_B T_e$, one can approximate $(\partial/\partial k_x) f_{even}(\vec{k})$ by a delta function centered around $k_F$. After integrating, and setting $k_F = \sqrt{\pi p}$ [61], the hole current in a 2-D graphene layer reads

$$I = W \frac{2q^2}{h} F v_f \tau(p) \sqrt{\pi p}, \tag{10}$$

where $W$ is the graphene layer width, $p$ is the hole concentration, $F$ is the electric field, and $\tau(p)$ is the relaxation time (inverse scattering rate) for a particular carrier concentration $p$. In the high field regime, one can assume $\tau(p) = \tau_{lf}/(1 + F/F_c)$, where $F_c$ is the critical electric field for the onset of high energy collisions such as remote phonons [62], for instance, $\tau_{lf}(p) = \tau_0 \sqrt{p/N_i}$ is the low-field relaxation time dominated by scattering with charged impurities with density $N_i$, and $\tau_0$ is a time constant. By setting $\mu_0 = (q/\hbar) v_F \tau_0 \sqrt{p/N_i}$, one recovers the conventional current expression

$$I = Wqpv(F), \tag{11}$$

with $v(F) = \mu_0 F/(1 + F/F_c)$, where the low-field conductance $\sigma_{lf} \propto p$, as observed experimentally [61].

## 2.3 Drift-diffusion carrier transport

In semiconductor physics, the drift-diffusion equation is related to drift current and drift velocity. The equation at the steady state for electrons and holes, respectively, is normally written as [20, 63]

$$\begin{cases} \frac{J_n}{-q} = -D_n \nabla_n - n\mu_n E \\ \frac{J_p}{q} = -D_p \nabla_p + p\mu_p E \end{cases}, \tag{12}$$

$$\begin{cases} \frac{\partial n}{\partial t} = -\nabla \frac{J_n}{-q} + R \\ \frac{\partial p}{\partial t} = -\nabla \frac{J_p}{-q} + R \end{cases}, \tag{13}$$

where $n$ and $p$ are the concentrations (densities) of electrons and holes, respectively,

$J_n$ and $J_p$ are the electric currents due to electrons and holes respectively, $\frac{J_n}{-q}$ and $\frac{J_p}{-q}$ are the corresponding "particle currents" of electrons and holes respectively, $R$ represents carrier generation and recombination, $E$ is the electric field vector, $\mu_n$ and $\mu_p$ are electron and hole mobility, respectively.

The diffusion coefficient and mobility are related by the Einstein relation following as [64]

$$\begin{cases} D_n = \mu_n k_B T/q \\ D_p = \mu_p k_B T/q \end{cases}, \tag{14}$$

The drift and diffusion current refer separately to the two terms in the expressions for $J$

$$\begin{cases} \frac{J_{drift}}{-q} = -n\mu E \\ \frac{J_{diffusion}}{-q} = -n\mu E \end{cases}, \tag{15}$$

where $\mu$ is the mobility ($\mu_n(\mu_p)$) of electrons (holes).

## 2.4 Density of states (DOS)

The density of states for monolayer graphene is expressed as [65]

$$\text{DOS}(E) = 2|E|/\pi(\hbar v_F)^2, \tag{16}$$

And the Fermi level to vary linearly with the voltage drop (i.e. $V_{CH}$) across the quantum capacitance $C_q$, which implies that $E_F = qV_{CH}$.

The DOS for bilayer graphene can be written as [66]

$$\text{DOS}(E) = \frac{2m_{eff}}{\pi\hbar^2}\left[H\left(E - \frac{E_g}{2}\right) + H\left(-E - \frac{E_g}{2}\right)\right], \tag{17}$$

where $m_{eff} = A$ is the effective mass, $E_g$ is the bilayer energy bandgap, $\hbar$ is the reduced Planck constant, H is the Heaviside step function, and A is a fitting parameter.

## 2.5 Surface potential

Surface potential, $\varphi_s$, i.e. the potential at the interface is an implicit function of the terminal voltages that is usually obtained by solving the surface potential equation (SPE) occasionally known also as the input voltage equation [67]. It is derived under several simplifying assumptions essential in the compact model formulation. The first simplification is the Shockley's gradual channel approximation (GCA) that assumes

$$\left|\frac{\partial^2 \varphi}{\partial y^2}\right| \ll \left|\frac{\partial^2 \varphi}{\partial x^2}\right|, \qquad (18)$$

where $\varphi$ denotes the electrostatic potential. With this simplification the Poisson equation becomes

$$\frac{\partial^2 \varphi}{\partial x^2} = -\frac{\rho}{\varepsilon_s}, \qquad (19)$$

where $\rho$ denotes the charge density and $\varepsilon_s$ is the dielectric permittivity. Denoting the electron and hole concentrations as $n$ and $p$, respectively,

$$\rho = q(p - n - N_a^-), \qquad (20)$$

where $N_a^-$ is the concentration of ionized acceptors and one can consider an $n$-channel MOS device. Since the hole current component is negligible, so it is the gradient of the hole imref $F_p$ and the hole concentration is given by the Boltzmann relation $p = p_b exp(-\beta\varphi)$ where $\beta = 1/\phi_t$ and $\phi_t$ denotes the thermal potential. This form assumes that the reference point for the potential is in the neutral bulk region where the majority and minority carrier concentrations are $p_b$ and $n_b$ respectively. For electrons it is necessary to take into account the imref gradient so that

$$n = n_b exp[\beta(\varphi - \phi_n)], \qquad (21)$$

where the normalized imref splitting (a.k.a. "channel voltage")

$$\phi_n = (1/q)(F_p - F_n). \qquad (22)$$

For most applications it is sufficient to assume the complete ionization of the channel dopants. Then $N_a^-$ coincides with the total acceptor concentration $N_a = p_b - n_b$ (for the uniformly doped channel). Hence

$$\rho = q[p_b(exp(-\beta\varphi) - 1) - n_b(kexp(-\beta\varphi) - 1)], \qquad (23)$$

where $k = exp(-\beta\phi_n)$.

From Eq. (19) and the boundary condition $\frac{\partial \varphi}{\partial x} = 0$ for $\varphi = 0$ it follows that

$$E_s^2 = -\frac{2}{\varepsilon_s}\int_0^{\varphi_s} \rho \, d\varphi, \qquad (24)$$

where the surface electric field $E_s = -(d\varphi/dx)_{x=0}$. Continuity of the normal component of the displacement vector at the Si/SiO$_2$ interface provides SPE in the form

$$\left(V_{gb} - V_{fb} - \varphi_s\right)^2 = \gamma^2 \phi_t h, \qquad (25)$$

where $V_{fb}$ is the flat-band voltage, $\gamma = \sqrt{2q\varepsilon_s p_b}/C_{ox}$ denotes the body factor, unit area oxide capacitance $C_{ox} = \varepsilon_{ox}/t_{ox}$, $t_{ox}$ is the oxide thickness and $\varepsilon_{ox}$ is the oxide permittivity. The dimensionless variable $h$ represents the normalized square of the surface electric field:

$$h = \frac{\varepsilon_s E_s^2}{2q\phi_t p_b} = -\frac{1}{q\phi_t p_b} \int_0^{\varphi_s} \rho \, d\varphi. \qquad (26)$$

## 3 Charge sheet compact model

The first compact model for GFET was proposed by Meric et al. based on charge sheet model [34, 35]. In Meric's model, it is assumed that a top-gated graphene FET is based on a high-$k$ gate dielectric without bandgap engineering. And the GFETs have source and drain regions that are electrostatically doped by the back gate, which enables control over the contact resistance and threshold voltage of the top-gated channel. Then, the sheet carrier concentrations (electrons or holes) in the source and drain regions are given as

$$n \cong \sqrt{n_0^2 + \left(C_{back}(V_{gs-back} - V_{gs-back}^0)/q\right)^2}, \qquad (27)$$

where $V_{gs-back}^0$ and $V_{gs-back}$ are the back-gate-to-source voltage and back gate voltage at the Dirac point in these regions, respectively, and $n_0$ is the minimum sheet carrier concentration as determined by disorder and thermal excitation. $C_{back}$ is the back-gate capacitance. Under the top gate, carrier concentrations are determined by both the front and back gates

$$n = \sqrt{n_0^2 + \left(C_{top}(V_{gs-top} - V_0)\right)^2}, \qquad (28)$$

where $V_0$, which has the character of a threshold voltage, is given by $V_0 \cong V_{gs-top}^0 + \frac{C_{back}}{C_{top}}\left(V_{gs-back}^0 - V_{gs-back}\right)$. $V_{gs-top}^0$ is top-gate-to-source voltage, $C_{top}$ is the top-gate capacitance and given by the parallel combination of the electrostatic capacitance of the gate and the quantum capacitance of graphene.

Then, the carrier concentration dependence of the distance in the channel, shown schematically in Fig. 3 for different points in the I-V trace, is calculated using a field-effect model:

$$n(x) = \sqrt{n_0^2 + \left(C_{top}(V_{gs-top} - V(x) - V_0)/q\right)^2}. \tag{29}$$

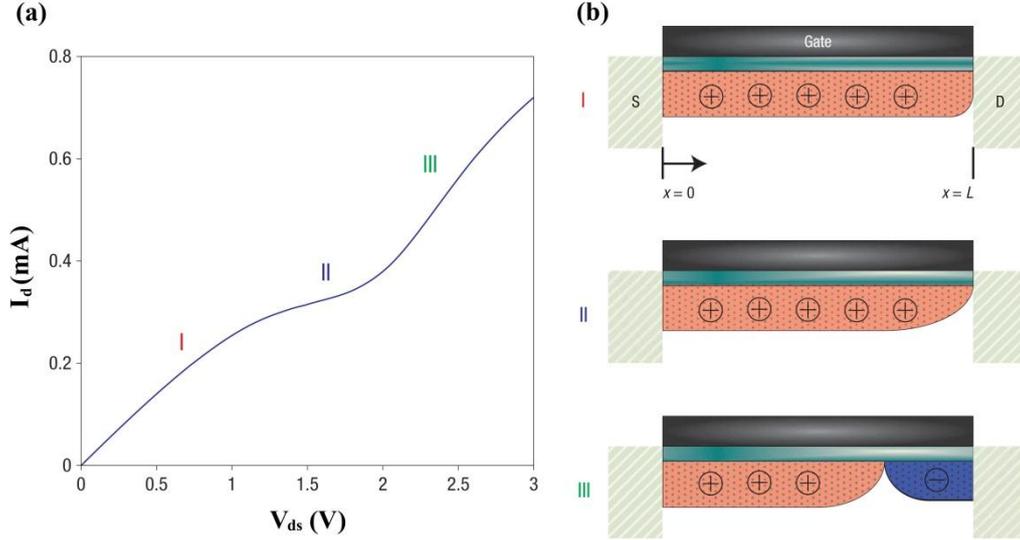

Fig. 3 (a) Measured I-V characteristic at $V_{gs-back}$=40 V and $V_{gs-top}$=0.3 V in GFET devices. Three points (I, II and III) are indentified in the I-V curve. (b) Schematic demonstration of the carrier concentration under the top-gated region. At point I ($V_{sd}$ , $V_{sd-kink}$, here $V_{sd-kink}$ means the $V_{sd}$ of "kink" between I and II, or II and III in Fig. 3(a)), the channel charge at the drain end begins to decrease as the minimal density point enters the channel. At point II ($V_{sd} < V_{sd-kink}$), the minimal density point forms at the drain. For $V_{sd} > V_{sd-kink}$ (point III), an electron channel forms at the drain.

With this consideration, the current in the channel is express by [68]

$$I_d = \frac{W}{L} \int_0^L qn(x)v_{drift}(x)dx, \tag{30}$$

where $L$ is the channel length and $W$ is the channel width. Current continuity forces a self-consistent solution for the potential $V(x)$ along the channel. Here approximating the carrier drift velocity ($v_{drift}$) by a velocity saturation model [69]

$$v_{drift}(x) = \frac{\mu E}{1 + \mu E/v_{sat}}, \tag{31}$$

where $v_{sat}$ is the saturation velocity of the carriers.

Combining Eqs. (29-31), the final current in the channel can be obtained as

$$I_d = \frac{\frac{W}{L}q\mu \int_0^{V_{sd}} \sqrt{n_0^2 + (C_{top}(V_{gs-top} - V(x) - V_0)/q)^2}}{1 + \frac{\mu V_{sd}}{Lv_{sat}}}, \tag{32}$$

Fig. 4 shows the results of simple field-effect modeling of the devices, compared with the measured I-V characteristics of GFET. This field-effect model is also implemented in Verilog-A with the equations. Here, $R_s$ is approximately 100 Ω (700 Ω) for two kinds of GFET devices, i.e., GFET A and GFET B. GFET A has a low-field mobility μ=$10^5$ cm²/Vs and $n_0$=5×$10^{12}$ cm⁻², compared with μ=1200 cm²/Vs and $n_0$=0.5×$10^{12}$ cm⁻² for GFET B.

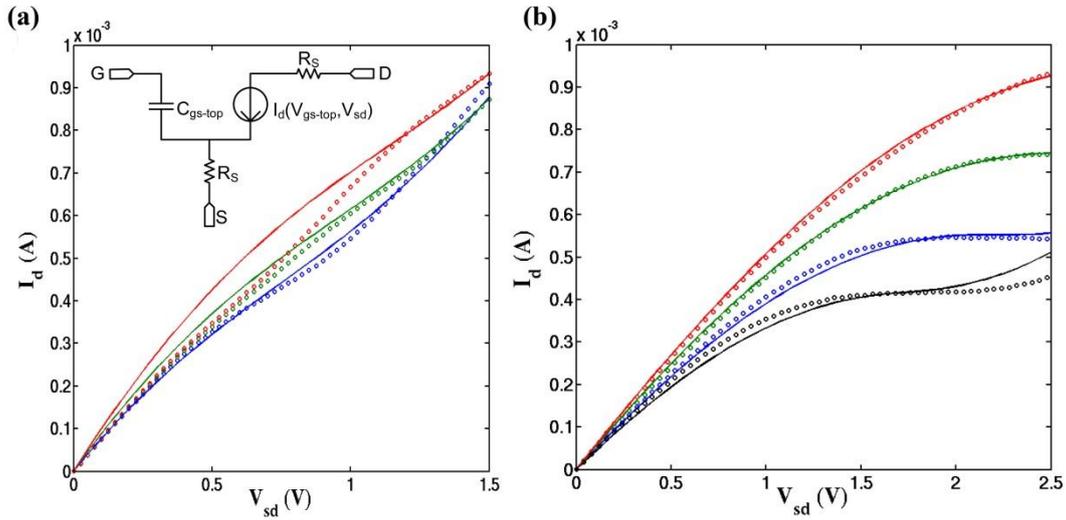

Fig. 4 Current-voltage characteristics of GFET device. Drain current ($I_d$) as a function of source-to-drain voltage ($V_{sd}$), (a) for GFET A at $V_{gs\text{-}top}$ = -0.1V, -0.3V, -0.5V and (b) for GFET B at $V_{gs\text{-}top}$ = 0V, -0.5V, -1.5V, -2.5V (from bottom to top), respectively. Both measured (solid curves) and simulated (dashed curves) results are shown. Inset in (a), Equivalent circuit for corresponding compact model of graphene FET.

Based on the Meric's model, Jaehong Lee et al. have developed a compact model of extremely scaled graphene FETs [38]. In Lee's model, an electron-hole puddle existing near the charge-neutral region (Dirac point) is considered at a low carrier density while the velocity saturation effect due to surface polar phonon scattering is included at a high carrier density. Otherwise, Sébastien Frégonèse et al. also proposed an electrical compact model for graphene FET device [39], at which, a trap model is introduced and the equivalent circuit is improved. And traps have an effect on the transconductance and influence consequently most figures of merit in circuit design. The electrical compact mode has been verified by comparison to DC and AC measurements versus bias and frequency on an advanced GFET having a transit frequency of about 10 GHz.

Although Meric's model has firstly been developed and shown better agreement between experimental data and simulated results of drain current, due to the difference between MOSFET and graphene FETs (GFETs), it maybe not entirely practicable to GFETs.

## 4 Compact model based on Boltzmann equation

To build a compact model accurately reflecting the characteristics of GFETs, researchers have presented the model based on Boltzmann equation [46]. Here, the graphene monolayer was assumed to sit on a thick $SiO_2$ layer with capacitance $C_{back}$ on top of a back gate that controls the source and drain resistance $R_s$, at the same time, as the channel threshold voltage with bias $V_{gback}$. Otherwise, a top gate of length L, separated from the graphene monolayer by a thinner oxide with capacitance $C_{top}$, controls the carriers in the channel with $V_{gtop}$. According to Boltzmann equation, by integrating the current of Eq. 11 from source to drain as in conventional MOS devices, and by taking into account the series resistance $R_s$ at the source and drain [35], Brett W. Scott et al. have obtained the drain current as [46]

$$I_d = \frac{W\mu_0 V_c}{2LC_{top}(|V_{ds}|-2|I_d|R_s+V_c)}[Q(L)^2 - Q(0)^2], \qquad (33)$$

where $Q(L) = -C_{top}(V_{g0} - V_{ds} - |I_d|R_s)$ and $Q(0) = -C_{top}(V_{g0} + |I_d|R_s)$, $V_{g0}$ is threshold voltage, here $V_{g0} = V_{gtop} - V_0$ and $V_0 = V_{top}^0 \frac{C_{back}}{C_{top}}(V_{gback}^0 - V_{gback})$, respectively, $R_s$ is the resistance. By solving for $I_d$, a closed expression for the drain current is expressed as

$$I_d = \frac{1}{4R_s}\left[V_{ds} - V_c + I_0 R_s + \sqrt{(V_{ds} - V_c + I_0 R_s)^2 - 4I_0 R_s V_{ds}}\right], \quad (34)$$

where $V_{ds}$ is the drain-source voltage, $I_0 = 2(W/L)\mu_0 V_c C_{top}(V_0 - V_{ds}/2)$, and $V_c = F_c L$, $F_c$ is the critical field.

Here, the low drain-source bias conductance is readily calculated by taking the derivative of the current expression (Eq. (34)) with respect to $V_{ds}$ as $V_{ds}$ goes to zero. One gets

$$g_{ds}(V_{ds} \to 0) = \frac{-V_{g0}}{|2R_s V_{g0} - R_c V_c|}, \qquad (35)$$

where $\frac{1}{R_c} = (W/L)\mu_0 C_{top} V_c$, so that $R_c V_c$ is independent of $V_c$, as is the conductance at low drain bias. The low drainsource bias resistance reads

$$R_{ds} = \frac{1}{g_{ds}} = 2R_s - \frac{R_c V_c}{V_{g0}}, \qquad (36)$$

which establishes a linear relation between $1/g_{ds}$ and $1/V_{g0}$ with a slope given by $R_c V_c$ (inversely proportional to the mobility) and an asymptotic conductance value for large $V_{g0}$ reaching $2R_s$.

In the same context, one obtains the expression for the drain-source saturation voltage as a function of the top gate voltage $V_{g0}$ by solving for $V_{ds}$ after setting the derivative of the current from Eq. (34) with respect to $V_{ds}$ equal to zero that yields

$$V_{ds(sat)} = \frac{2\gamma V_{g0}}{1+\gamma} + \frac{1-\gamma}{(1+\gamma)^2}\left[V_c - \sqrt{V_c^2 - 2(1+\gamma)V_c V_{g0}}\right], \qquad (37)$$

with $\gamma = R_s/R_c$. Substituting Eq. (37) into Eq. (34), one can obtain the expression of the saturation drain current as a function of the top gate voltage that reads

$$I_{d(sat)} = \frac{\gamma}{R_s(1+\gamma)^2} \times \left[-V_c + (1+\gamma)V_{g0} + \sqrt{V_c^2 - 2(1+\gamma)V_c V_{g0}}\right]. \qquad (38)$$

By taking the derivative of the saturation current with respect to the top gate voltage, one derives the expression for the transconductance at saturation

$$g_m^{sat} = \frac{1}{R_s + R_c}\left[1 - \frac{1}{\sqrt{1 - 2(1+\gamma)V_{g0}/V_c}}\right]. \qquad (39)$$

Fig. 5 shows the plots of both the low-bias conductance $g_{ds}$ as a function of the top gate voltage, and the low-bias resistance $R_{ds}$ as a function of the inverse of the top gate voltage in the device configuration investigated in Ref. [35]. In Fig. 5(a), the solid curve is calculated from Eq. (35), which shows good agreement with the experimental data close to the minimum conductance, but underestimates the former by about 20% at high top gate bias. In Fig. 5(b), the experimental resistance values display a linear relation with $1/V_{g0}$ in agreement with Eq. (36).

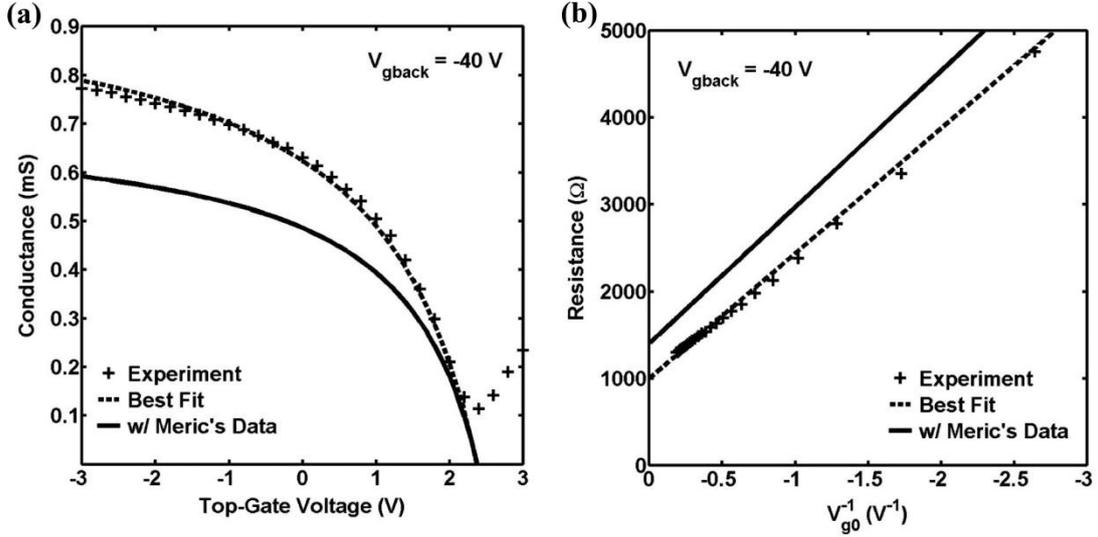

Fig. 5 (a) Small-signal source-drain conductance $g_{ds}$ as a function of the top gate voltage minus the threshold voltage $V_{g0}$, and (b) small-signal source-drain resistance $R_{ds}$ as a function of the inverse of the top gate voltage minus the threshold voltage $1/V_{g0}$.

In Fig. 6(a), the authors display the *I-V* characteristics of the GFET. An excellent agreement between the experiment and simulated results from Eq. (34) is obtained. However, the mobility is 25% higher than the Meric's fitted values [35]. Fig. 6(b) shows the comparison between theoretical and experimental results for both the p-channel conductance. The better fits are shown as compared with Meric's results [35]. Besides the simulated drain current, Brett W. Scott et al. in their compact model have predicted a linear dependence of the low-field resistance versus inverse gate voltage, and suggested that nonlinearity in the energy dispersion should be included, as well as carrier multiplication by impact ionization.

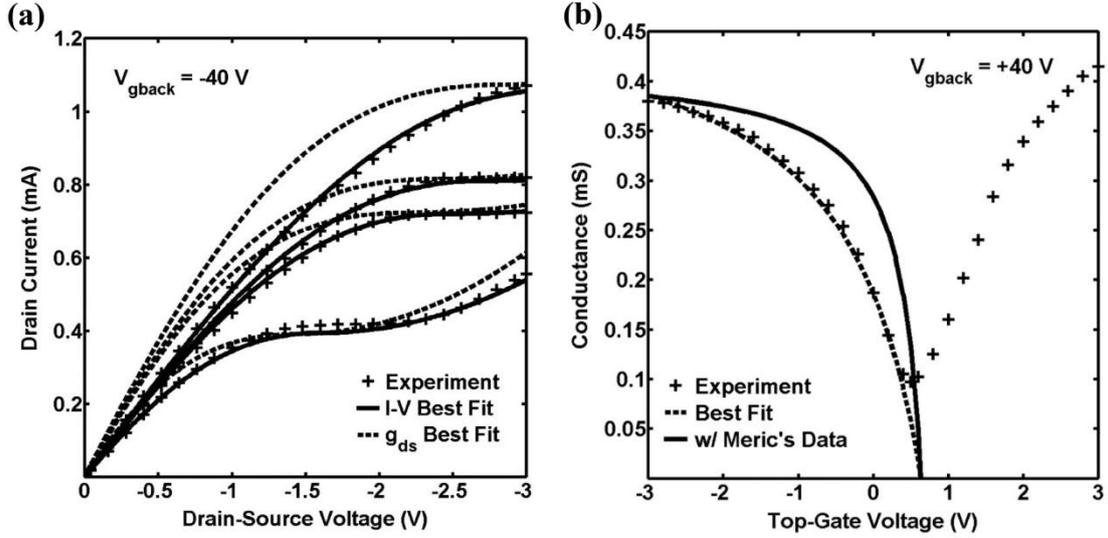

Fig. 6 (a) small-signal source-drain current $I_d$ as a function of drain-source voltage $V_{ds}$, and (b) small-signal source-drain conductance $g_{ds}$ as a function of the top gate voltage minus the threshold voltage $V_{g0}$ .

## 5 Compact model based on drift-diffusion carrier transport

Based on drift-diffusion carrier transport, David Jiménez et al. have presented a physics-based compact model of the current-voltage characteristics of GFET, of especial interest for analog and RF applications where band-gap engineering of graphene could be not needed [41, 42]. David Jiménez et al. assumed that the electrostatics of this device was the equivalent capacitive circuit. At which, the equivalent capacitive circuit $C_t$ and $C_b$ are the top and bottom oxide capacitance, and $C_q$ represents the quantum capacitance of the graphene. Potential $V_c$ represents the voltage drop across $C_q$, where $C_q = k|V_c|$ under condition $qV_c \gg k_BT$ , with $k = (2q^2/\pi)(q/(\hbar v_F)^2)$, and $v_F(= 10^6 \; m/s)$ is the Fermi velocity. Potential $V(x)$ is the voltage drop in the graphene channel, which is zero at the source end at $x$=0 and equal to drain-source voltage $V_{ds}$ at the drain end at $x$=L. Then, to model the drain current, a drift-diffusion carrier transport is assumed under form

$$I_{ds} = -qW\rho_c(x)v(x), \qquad (40)$$

where $W$ is the gate width, $\rho_c(x) = \frac{|Q_c|}{q}$ is the free carrier sheet density in the channel at position $x$, and $v(x)$ is the carrier drift velocity. Using a soft-saturation model, consistent with Monte Carlo simulations [70], $v(x)$ can be expressed as

$$v = \mu E/(1 + \mu|E|/v_{sat}), \qquad (41)$$

where $E$ is the electric field, $\mu$ is the carrier low field mobility, and $v_{sat}$ is the saturation velocity. The latter is concentration dependent and given by $v_{sat} = \Omega/\sqrt{\pi \rho_c}$. Applying $E = -dV(x)/dx$, combining the above expressions for $v$ and $v_{sat}$, and integrating the resulting equation over the device length, the drain current becomes

$$I_{ds} = \frac{q\mu W \int_0^{V_{ds}} \rho_c dV}{L + \mu \left| \int_0^{V_{ds}} \frac{1}{v_{sat}} dV \right|}. \tag{42}$$

The denominator represents an effective length ($L_{eff}$) to take into account the saturation velocity effect. In order to get an explicit expression for $I_{ds}$, the integral in Eq. (42) is solved using $V_c$ as the integration variable and consistently expressing $\rho_c$ and $v_{sat}$ as a function of $V_c$,

$$I_{ds} = \frac{q\mu W \int_{V_{cs}}^{V_{ds}} \rho_c(V_c) \frac{dV}{dV_c} dV_c}{L + \mu \left| \int_{V_{cs}}^{V_{ds}} \frac{1}{v_{sat}(V_c)} \frac{dV}{dV_c} dV_c \right|}, \tag{43}$$

where $V_c$ can be written as

$$V_c = \frac{-(C_t + C_b) + \sqrt{(C_t + C_b)^2 + 2k[(V_{gs} - V_{gs0} - V)C_t + (V_{bs} - V_{bs0} - V)C_b]}}{\pm k}. \tag{44}$$

The positive (negative) sign applies whenever $(V_{gs} - V_{gs0} - V)C_t + (V_{bs} - V_{bs0} - V)C_b > 0$. The channel potential at source $V_{cs}$ is determined by $V_c(V=0)$. Similarly, the channel potential at drain $V_{cd}$ is determined by $V_c(V=V_{ds})$. On the other hand, the charge sheet density can be written as $\rho_c(V_c) = kV_c^2/2q$. Extra term $\rho_0$ added to $\rho_c$ accounts for the carrier density induced by impurities [71]. Inserting these expressions into Eq. (42), the following explicit drain current expression can be finally obtained:

$$I_{ds} = \frac{\mu W}{L_{eff}} \left\{ q\rho_0 V_{ds} - \frac{k}{6}(V_{cd}^3 - V_{cs}^3) - \frac{k^2}{8(C_t + C_b)}(sgn(V_{cd})V_{cd}^4 - sgn(V_{cs})V_{cs}^4) \right\}. \tag{45}$$

To test the model, David Jiménez et al. have benchmarked the resulting I–V characteristics with experimental results extracted from devices in Refs. [35] and [72], as shown in Fig. 7. The first device under test has L=1 μm and W=2.1 μm, top dielectric is HfO$_2$ of 15 nm, and permittivity is ~16, and the bottom dielectric is silicon oxide of 285 nm. The back-gate voltage was −40 V. The flat-band voltages $V_{gs0}$ and $V_{bs0}$ were tuned to 1.45 and 2.7 V, respectively. These values were selected to locate the Dirac point according to the experiment. Fig. 8 shows the resulting I-V characteristics. Here the authors have extended the simulated voltage range beyond the experiment range to show the predictive behavior of the model. The transfer characteristics exhibit an ambipolar behavior dominated by holes (electrons) for $V_{gs} - V_{gs0} < V_{gs,D} (> V_{gs,D})$ where $V_{gs,D}$ (Dirac gate voltage) is given by $V_{gs,D} = V_{gs0} + V_{ds}/2$. The output characteristics behave is similar as the first examined device. Once again, the comparison between the model and experiment further demonstrates the accuracy of the model.

Based on drift-diffusion carrier transport, David Jiménez et al. have presented a physics-based compact model of current-voltage characteristics of GFETs, which can capture the physics of all operation regions within a single expression for the drain current and each terminal charge and capacitance. At the same time, it is of special interest as a tool for the design of analog and RF applications. Additional physical effects as, for example, short-channel effects, nonquasi-static effects, extrinsic capacitances, and mobility model, need to be incorporated into the long-channel core presented here to build a complete GFET compact model.

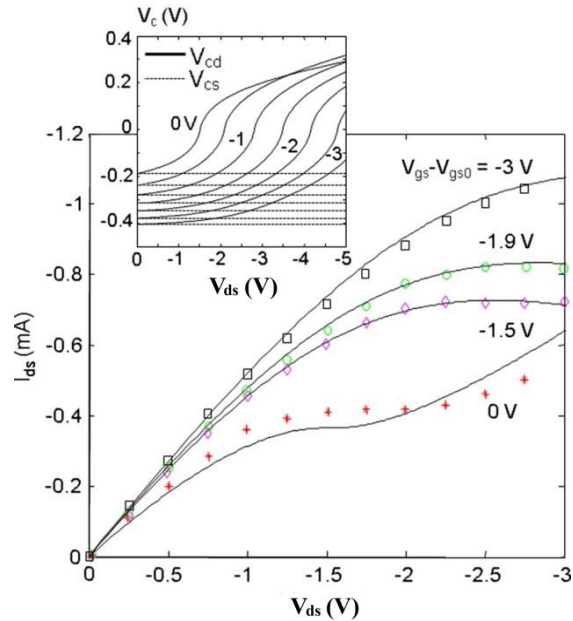

Fig. 7. Output characteristics obtained from the (solid lines) analytical model compared with experimental results from (symbols) [35]. Inset: quantum capacitance voltage drop as a function of the source-drain voltage for different gate voltage overdrive.

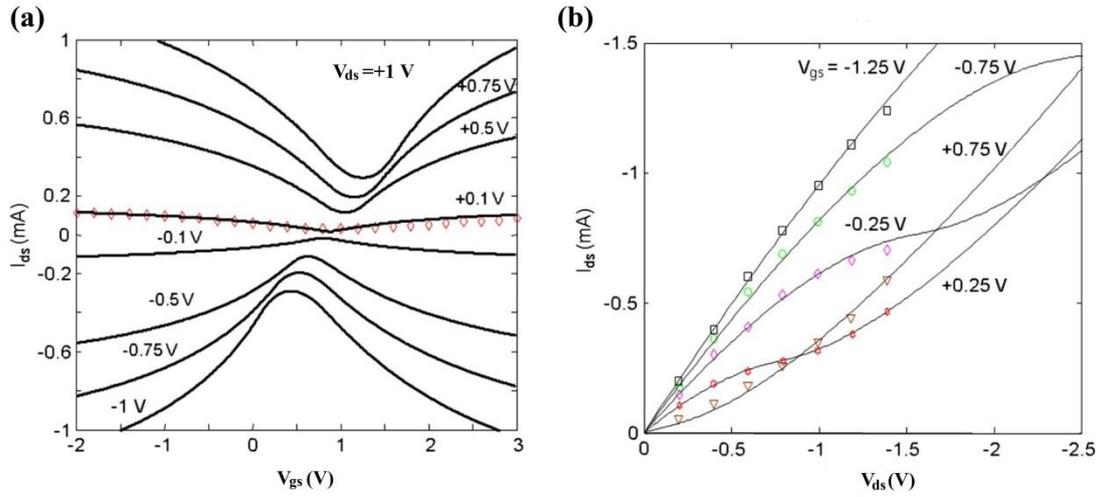

Fig. 8 (a) Transfer and (b) output characteristics obtained from the (solid lines) compact model compared with experimental results from (symbols) [72].

## 6 Compact model based on density of states (DOS)

To evaluate the capabilities of electronic devices based on graphene as a channel material, accurate models are desired. Some new compact models based on physics-based DOS have been also developed. For example, J. D. Aguirre-Morales et al. have developed a new compact model for monolayer graphene FETs and dual-gate bilayer GFET by using a physics-based DOS to simulate the current and the charge, respectively [32, 49-51].

### 6.1 Monolayer graphene FETs

For the monolayer graphene FETs, the model considering the 2-D DOS of monolayer graphene is used as Eq. (16). The corresponding device and equivalent capacitive circuit of monolayer GFET are shown in Fig. 9 [49]. As shown in Fig. 9(a), the graphene film is located between the top and the back-gate dielectrics. The source and drain ohmic contacts as well as the top-gate stack are located on the top of the graphene channel. The back-gate stack is comprised of a dielectric and a substrate acting as the back-gate. In this structure, the access region resistances are modeled as a function of the applied back-gate bias. The model is based on the 2-D DOS of monolayer graphene being proportional to the Fermi level energy, $E_F$.

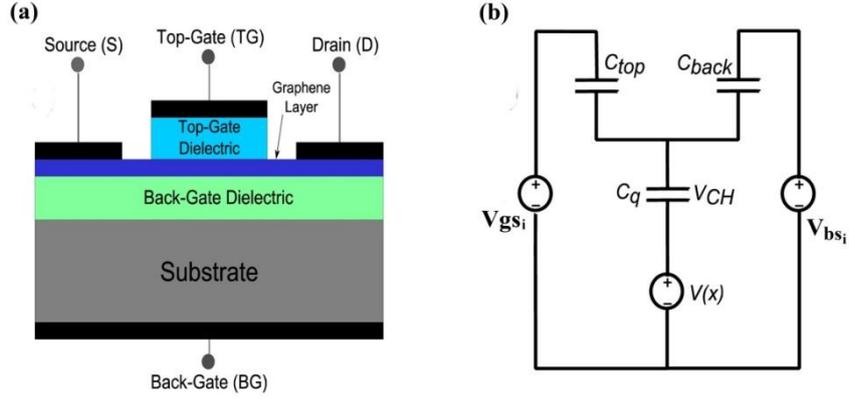

Fig. 9. (a) Cross section view and (b) equivalent capacitive circuit of monolayer GFET structure.

For the monolayer GFETs, the drain current has also been derived from the general drift-diffusion equation and considered the velocity saturation of carriers [73], was expressed as

$$I_{ds} = \mu W \frac{\int_0^{V_{ds_i}} (|Q_{net}| + q \cdot n_{puddle})}{L + \mu \left| \int_0^{V_{ds_i}} \frac{1}{v_{sat}} dV \right|}, \tag{46}$$

where $Q_{net}$ is the stored charge density in the channel and $n_{puddle} = \Delta^2/\pi(\hbar v_F)^2$ accounts for the formation of hole and electron puddles in the graphene sheet [66]. $\Delta$ represents the spatial inhomogeneity of the electrostatic potential and $v_F$ the Fermi velocity. To account for this asymmetric conduction behavior, a separation of electron and hole branch contributions is necessary for the total drain current implementation, the total drain current can be written as the sum of the hole and electron contributions ($I_{ds} = I_{ds_n} + I_{ds_p}$) as

$$I_{ds} = \mu_n W \frac{\int_0^{V_{ds_i}} (|Q_n| + \frac{q \cdot n_{puddle}}{2})}{L + \mu_n \left| \int_0^{V_{ds_i}} \frac{1}{v_{sat_n}} dV \right|} + \mu_p W \frac{\int_0^{V_{ds_i}} (|Q_p| + \frac{q \cdot n_{puddle}}{2})}{L + \mu_p \left| \int_0^{V_{ds_i}} \frac{1}{v_{sat_p}} dV \right|}, \tag{47}$$

where ($Q_n$, $v_{sat_n}$, $\mu_n$) and ($Q_p$, $v_{sat_p}$, $\mu_p$) are the charge sheet density, saturation velocity and mobility for electrons and holes, respectively. The net stored charge density in the channel can be written as

$$Q_{net}(x) = Q_p - Q_n = -\frac{1^3}{\pi(\hbar v_F)^2} V_{CH}(x) |V_{CH}(x)|. \tag{48}$$

Under the applied bias voltages ($V_{gs}, V_{ds}, V_{bs}$), one can write a second degree equation of the channel potential, $V_{CH}$, solving the charge equations from the equivalent capacitive circuit shown in Fig. 9(b) as

$$V_{CH}(x) = sign(Q_{tot} - C_{eq}V(x)) \frac{-C_{eq} + \sqrt{C_{eq}^2 + 4\alpha|Q_{tot} - C_{eq}V|}}{2\alpha}, \qquad (49)$$

with $C_{eq} = C_{top} + C_{back}$ , $Q_{tot} = C_{top}V_{gsi} + C_{back}V_{bsi} + qN_F$ and $\alpha = q^2/\pi(\hbar v_F)^2$, where $C_{top}$ and $C_{back}$ are the top and back gate capacitances, respectively. $N_F$ is the net doping in the graphene channel. $V_{gsi}$ and $V_{bsi}$ are the intrinsic voltages within the structure and $V(x)$ is the voltage drop across the graphene channel. $V_{CH}(x)$ along the channel depends on the bias conditions of the graphene layer. Then, based on Eq. (49), the net charge due to electron and hole conductions lead to the following equations, respectively

$$Q_{net} \approx \begin{cases} -Q_n \to -Q_n \approx \alpha V_{CH}|V_{CH}| = \alpha V_{CH}^2 \\ -Q_p \to -Q_p \approx \alpha V_{CH}|V_{CH}| = \alpha V_{CH}^2 \end{cases}. \qquad (50)$$

Finally, the gate-source and gate-drain capacitances have been calculated using the following equations

$$\begin{cases} C_{gs} = \left(\frac{dQ_{channel}}{dV_{gs}}\right)_{V_{ds}=cst} + \left(\frac{dQ_{channel}}{dV_{ds}}\right)_{V_{ds}=cst} \\ C_{gd} = \left(\frac{dQ_{channel}}{dV_{gd}}\right)_{V_{gs}=cst} = -\left(\frac{dQ_{channel}}{dV_{ds}}\right)_{V_{gs}=cst} \end{cases}, \qquad (51)$$

with $Q_{channel} = W \int_0^L (Q_{net} + q \cdot n_{puddle}) \, dx$.

For verifying the model as described above, measurement results from two different monolayer GFET technologies have been compared with the model simulations. Figure 10 shows the DC characteristics ($I_d$-$V_{gs}$, $I_d$-$V_{bs}$, $I_d$-$V_{ds}$) from a device having a gate length of 5 μm and width of 25 μm [36]. The simulation shows a good agreement with measured results.

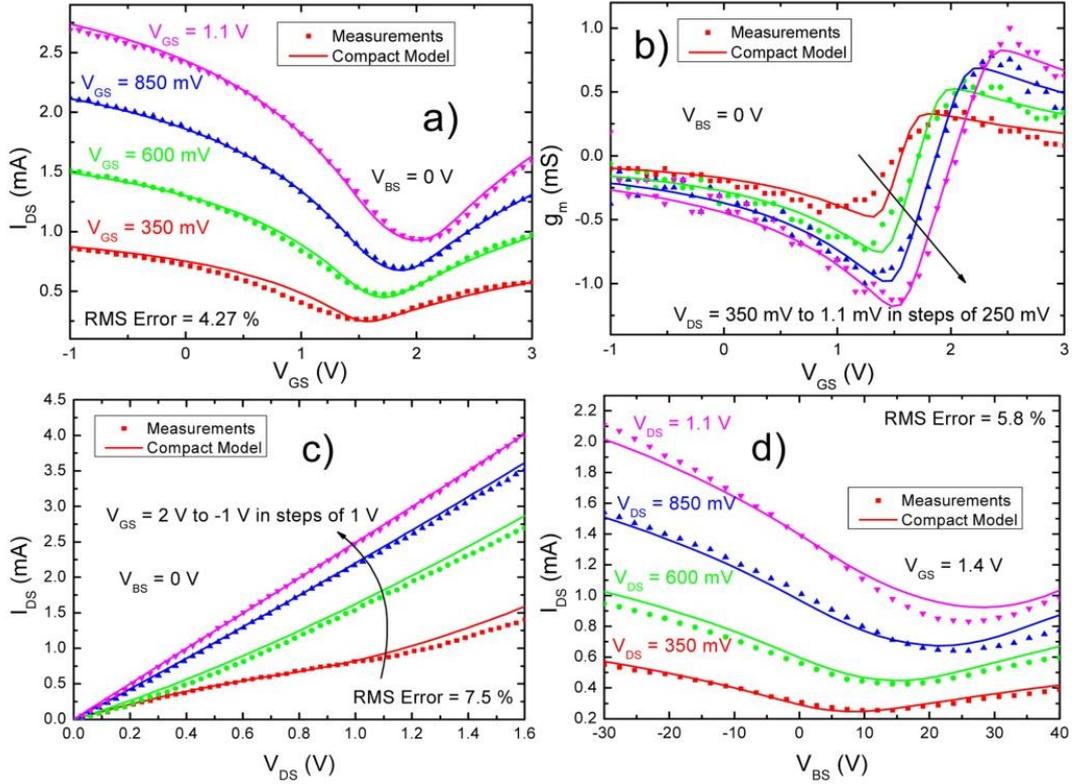

Fig. 10 Comparison of (a) $I_d$-$V_{gs}$, (b) $g_m$-$V_{gs}$, (c) $I_d$-$V_{ds}$ and (d) $I_d$-$V_{bs}$ measurements (symbols) with the compact model (solid lines).

## 6.2 Bilayer graphene FETs

For the bilayer graphene FETs, the model considering DOS of bilayer graphene is used as Eq. (17). The corresponding device and equivalent capacitive circuit of bilayer GFET are shown in Fig. 11 [49]. It is assumed that on top of the bilayer graphene channel, the source and drain ohmic contacts as well as the top-gate stack are located. The back-gate stack is composed of a dielectric and a substrate acting as the back-gate.

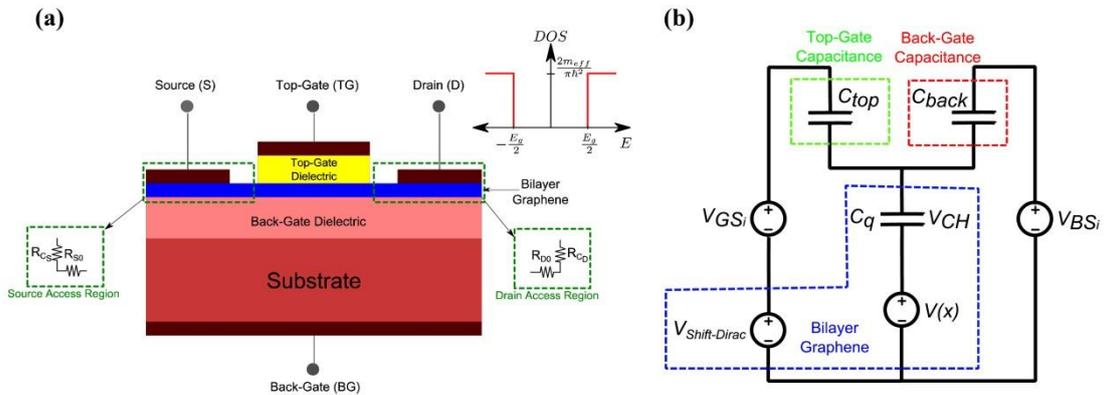

Fig. 11 (a) Cross-sectional view and the equivalent drain and source access resistances,

and (b) equivalent capacitive circuit of bilayer GFET structure. Inset: DOS of bilayer graphene with a nonzero energy bandgap.

The drain current $I_{ds}$ in the bilayer graphene FETs also can be expressed as the sum of the electron and hole contribution, $I_{ds} = I_{ds_n} + I_{ds_p}$, which is similar as that in the monolayer graphene FETs.

The saturation velocity of carriers is given by

$$V_{sat_{(n,p)}} = \frac{\hbar\Omega}{\sqrt{\pi\left(\frac{|Q_{(n,p)}|}{q} + \frac{n_{puddle}}{2}\right)}}, \tag{52}$$

with $\hbar\Omega$ being the effective energy due to phonon scattering in the substrate. Here, an even distribution of the residual carrier density, $n_{puddle}$, in the electron and hole puddles is considered.

As the back-gate voltage, $V_{bs}$, becomes more negative, more positive charges are induced to close the back-gate, thereby inducing more negative charges close to the graphene channel on the top side of the back-gate dielectric. Thus, higher values of $V_{gs}$ are required to obtain the channel charge inversion, resulting in a shift in the Dirac point, $V_{Shift-Dirac}$, toward the positive direction. Then, the shift in the Dirac voltage will also saturate eventually as described by the exponential model

$$V_{Shift-Dirac} = V_1 + V_2 \cdot exp\left(-\frac{|V_{bs_i}|}{V_3}\right), \tag{53}$$

where $V_1$, $V_2$, and $V_3$ are constants.

Otherwise, considering that the area of hole and electron puddles are equal in size, the spatial electrostatic potential is simplified as a step function with a peak-to-peak value of $\pm\Delta$, as presented in Ref. [66], which includes the effect of the opening of an energy bandgap. The electron and hole puddles are written as

$$n_{puddle} = \frac{2m_{eff}}{\pi\hbar^2}k_BT\left\{\ln\left[1 + exp\left(\frac{-\frac{E_g}{2}+\Delta}{k_BT}\right)\right] + \ln\left[1 + exp\left(\frac{\frac{E_g}{2}-\Delta}{k_BT}\right)\right]\right\}. \tag{54}$$

In order to validate the capability of the developed compact model for accurate modeling of the bilayer GFET devices, comparisons with measurements from the literature [74] have been performed and are presented. The comparison of the measurement data and the developed compact model is shown in Figs. 12, which shows the good agreement.

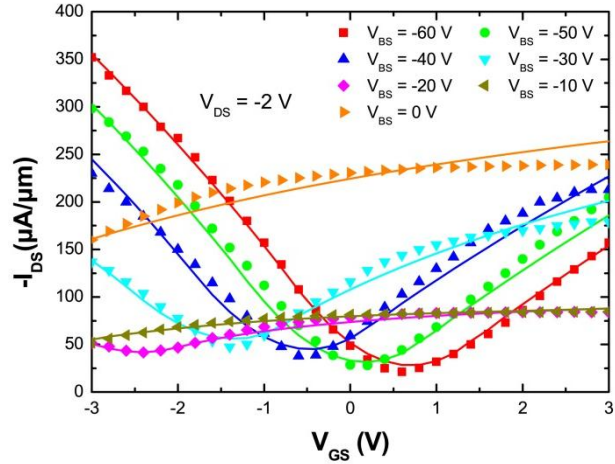

Fig. 12 Comparison of the $I_{ds}$–$V_{gs}$ measurements [34] (symbols) with the compact model (solid lines) for $V_{ds}$=−2 V and $V_{bs}$=0 to −60 V.

## 7 Surface-potential-based compact model

As described above, plenty of compact models for GFETs have been developed, with the current characteristics and can be implemented in Verilog-A (VA) for simulations of DC, AC or transient properties. However, to achieve an accurate model qualified in EDA tool, the compact model with high accuracy and strong physical property should be further considered. For the real graphene device, the random distribution of carriers (electron or hole) caused by the impurities and spatial disorder, will play a role in carrier transport. Thus, to obtain an accurate transport property, compact model in GFETs should include the real physical effect, such as, disorder and temperature. The surface- potential-based compact model is believed to have high accuracy and strong physical property, and be easily simplified into the charge-based and threshold-voltage-based model.

Based on the surface-potential-based, Li et al. have developed a continuous physical compact model for GFET. Fig. 13(a) shows the schematic diagram of a top-gated GFET for GEFT compact model [47, 48]. In Li et al.'s work, it is assumed that on top of the graphene channel, the source and drain ohmic contacts as well as the top-gate stack are located. Other, DOS for graphene is Gaussian DOS with disorder parameter, and the analytical carrier density is based on the exponential distribution of potential fluctuations for the electron branch ($E_F > 0$) or the hole branch ($E_F < 0$).

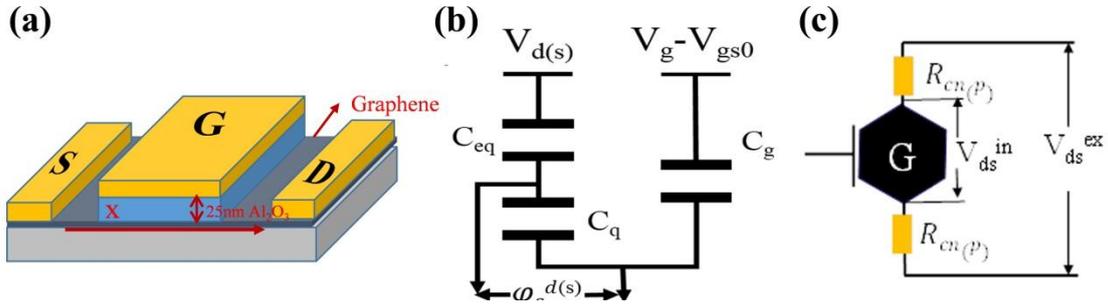

Fig. 13 (a) Geometric definition for compact model of GFET, (b) capacitance divider scheme extracted from a single gated GFET, and (c) extrinsic structure for GFET

## 7.1 Surface-potential- based model

Generally, the threshold voltage emphasizes the converting point, while the surface potential, $\varphi_s$, always generates a continuous transition behaviors covering the whole region. For the zero band-gap graphene, carriers generally degenerate with Fermi level located in the conduction or valence energy band. In this situation, the surface potential is obtained by using capacitance divider scheme instead of Poisson equation as shown in Fig. 13(b). For a single gate GFET, $\varphi_s$ is expressed as

$$\varphi_s = sgn(V_x)\frac{-C_g + \sqrt{C_g^2 + 4kC_gV_{gx}}}{2k}, \tag{55}$$

where $sgn$ is the sign function, $V_{gx} = \sqrt{\left(V_{gs} - V_{gs}^0 - V(x)\right)^2}$, $V(x)$ is the voltage along the channel, $V(0)$ or $V(L)$ represents the source or drain voltage, $V_{gs}^0$ is the gate doping voltage which will cause the drift of Dirac point. Eq. (55) is applicable for high gate voltage region arising from the simple quantum capacitance. However, at the low gate voltage, the disorder effect dominates the transport and hence demands the accurate $C_q$. According to the Gaussian distribution of potential fluctuations [75], although $C_q$ can achieve the high accuracy, it also blocks the analytical solution of $\varphi_s$. Fortunately, this question has been addressed by Li $et$ $al$. by using the exponential distribution instead of Gaussian form [76].

Based on Ref. [76], the surface potential $\varphi_s$ can be obtained by solving the following equation,

$$V_{gx} = q(n_e(\varphi_s) - n_h(\varphi_s)) \left( \frac{1}{C_g} + \frac{1}{C_{eq}} \right) + \varphi_s, \tag{56}$$

where $C_{eq}$ is the effective capacitance of the metal and graphene contact, $C_{eq} = \varepsilon_0/d_{eq}$ and $d_{eq}$ is the effective contact distance. By using $3^{rd}$ Taylor series, the final surface potential $\varphi_s$ can be written as

$$\begin{cases} f_4 = \sqrt{(27f_1^2 + 9f_1f_2f_3 - 2f_2^3)^2 + 4(3f_1f_3 - f_2^2)^3} + 27f_1^2 + 9f_1f_2f_3 - 2f_2^3 \\ \varphi_s = \frac{\sqrt[3]{f_4}}{3\sqrt[3]{2}f_1} - \frac{\sqrt[3]{2}(3f_1f_3 - f_2^2)}{3f_1\sqrt[3]{f_4}} - \frac{f_2^2}{3f_1} \end{cases}, \tag{57}$$

## 7.2 Asymmetry Behaviors

It is well known that ambipolar characteristic curves in GFET show an asymmetry for electron and hole, which mainly derived from contact metal doping. Thus, the corresponding model should be modified. Based on the contact resistances as shown in Fig. 13(c), Li et al. developed the intrinsic voltage as [47, 48]

$$V_{ds}^{in} = V_{ds}^{ex} - 2I_{ds}R_c, \tag{58}$$

where $V_{ds}^{in}$ and $V_{ds}^{ex}$ are the intrinsic and extrinsic drain-source voltage, respectively, $I_{ds}$ is the drain-source current related to $V_{ds}^{in}$. Eq. (58) can only be solved numerically with contact resistance for hole (electron). Considering the contribution of the contact doping to the surface potential, the asymmetry surface potential can be obtained by altering the effective contact capacitance ($C_{eq}$) in different operation regions. Then, the asymmetry surface potential is written as

$$\varphi_{sa} = \frac{1}{r_0} log[exp(r_0 sgn(V_{gx})\varphi_{sh}) + exp(r_0 sgn(V_{gx})\varphi_{se})], \tag{59}$$

where $r_0$ is a fitting parameter and its sign controls the type of metal doping, the detailed parameter is shown in Table I.

Based on Eq. (59), one can calculate the surface potential with numerical and analytical solutions, as shown in Fig. 14. The corresponding small error is shown in inset.

Table I Different metal-graphene contact and channel related to fitting parameter $r_0$.

| Metal | Contact form | Parameter |
|-------|-------------|-----------|
| Ag (Ti, Au, Pd) | p-p-p | $r_0 < 0,\ V_{gs} < V_{dirac}$ |
| | p-n-p | $r_0 < 0,\ V_{gs} > V_{dirac}$ |
| Al | n-p-n | $r_0 > 0,\ V_{gs} > V_{dirac}$ |

| n-n-n | $r_0 > 0,\ V_{gs} > V_{dirac}$ |
|---|---|

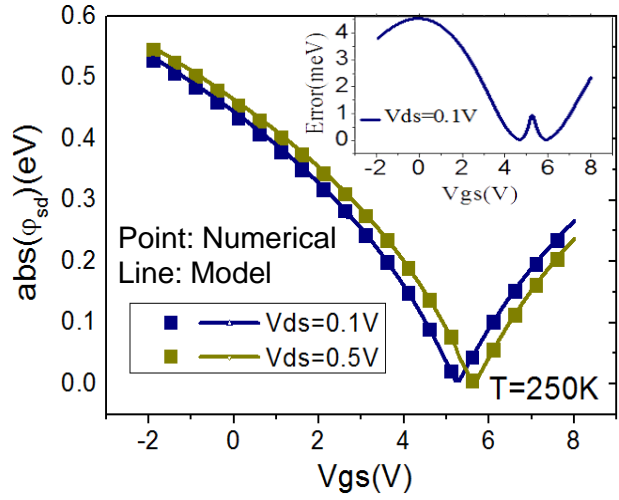

Fig. 14 Simulation results of surface potential with numerical and analytical solutions, respectively. Inset: an error between numerical and analytical solutions.

### 7.3 Current modeling

Generally, the constant mobility instead of function of Fermi level in model was applied. To achieve a higher level, the functional current will be simplified and then reduce accuracy. To obtain achieve carrier transport property, Li *et al*. introduced thermal activated transport theory and Bolzmann transport theory into the surface-potential-based model [48]. Electron or hole will contribute to the thermal activated transport [77] in the low Fermi level region, while in the high Fermi level the Boltzmann transport will be dominant and include two scattering mechanisms (long-range and short-range). Based on these theories, the current should include three components as

$$
\begin{cases}
I_{long} = \frac{W}{L} \frac{qkh\pi}{l_0 n_i \left(\frac{2\pi q^2}{k}\right)^2} \left[ sgn(\varphi_s) \int (n_e(\varphi_s) + n_h(\varphi_s)) \varphi_s \frac{dV(x)}{d\varphi_s} d\varphi_s \right] \Big|_{|\varphi_{ss}|}^{|\varphi_{sd}|} \\
I_d = \frac{W}{L} \left[ sgn(\varphi_s) \int \sigma_d \frac{dV(x)}{d\varphi_s} d\varphi_s \right] \Big|_{|\varphi_{ss}|}^{|\varphi_{sd}|} \\
I_{short} = \frac{W}{L} \frac{4kD_1 h(hV_f)^2}{q^3 n_d V_0^2} \left[ sng(\varphi_s) \left( \int \left( 1 - exp\left( -\frac{\varphi_s}{\beta} \right) \right) \frac{dV(x)}{d\varphi_s} d\varphi_s \right) \right] \Big|_{|\varphi_{ss}|}^{|\varphi_{sd}|}
\end{cases}
, \quad (60)
$$

where the subscripts *d*, *long* and *short* represent the thermal activated, long range scattering and short-range scattering mechanisms, respectively, $\varphi_{sd}$ ($\varphi_{ss}$) represents the drain (source) surface potential obtained by Eq. (57). Due to the competence of the two scattering mechanisms and the geometric mean presenting the transition from the thermal activation to Boltzmann transport, the final current model is expressed as

$$I_{total} = \left[ \left( I_{long}^{-r_1} + I_{short}^{-r_1} \right)^{-r_2/r_1} + I_d^{r_2} \right]^{1/r_2}, \qquad (61)$$

where $r_1$ and $r_2$ are the fitting parameters.

In Fig. 15(a) and Fig. 15(b), the comparison of the calculated and experimental results for output and transfer current characteristics is presented.

For current models, the output current saturation is always considered, which arises from drift velocity saturation or Joule heating effects. Generally, model with drift velocity saturation is written as the effective length expression,

$$L_{eff} = L + \int_{\varphi_{ss}}^{\varphi_{sd}} \frac{\mu}{v_{sat}(\varphi_s)} \frac{dV(x)}{d\varphi_s} d\varphi_s, \qquad (62)$$

where $v_{sat}$ is the saturation velocity. As well as typical models for $v_{sat}$ and μ, various effective channel length expressions can be obtained. Actually Eq. (62) can be fully simplified with constant mobility and non-disorder carrier density (n∝$\varphi_s^2$). Then, Eq. (62) will be transformed as,

$$L_{eff} - L \propto \left( \varphi_s^2 + C\varphi_s^3 \right) \Big|_{\varphi_{ss}}^{\varphi_{sd}}. \qquad (63)$$

According to Eq. (63), model including residue carrier density ($n_0$) or disorder parameter (s) has been evolved. Remarkably, the effective length with the scattering mechanisms dependent mobility can be improved. $L_{eff}$ embedded with two scattering mechanisms can be simply presented as,

$$I_{eff} - L \propto \left( L_{long}^{-r_1} + L_{short}^{-r_1} \right)^{-\frac{1}{r_1}}. \qquad (64)$$

Based on Eq. (64), the comparison of saturation output current for the calculated and experimental results under different channel lengths is presented in Fig. 15(c) and Fig. 15(d).

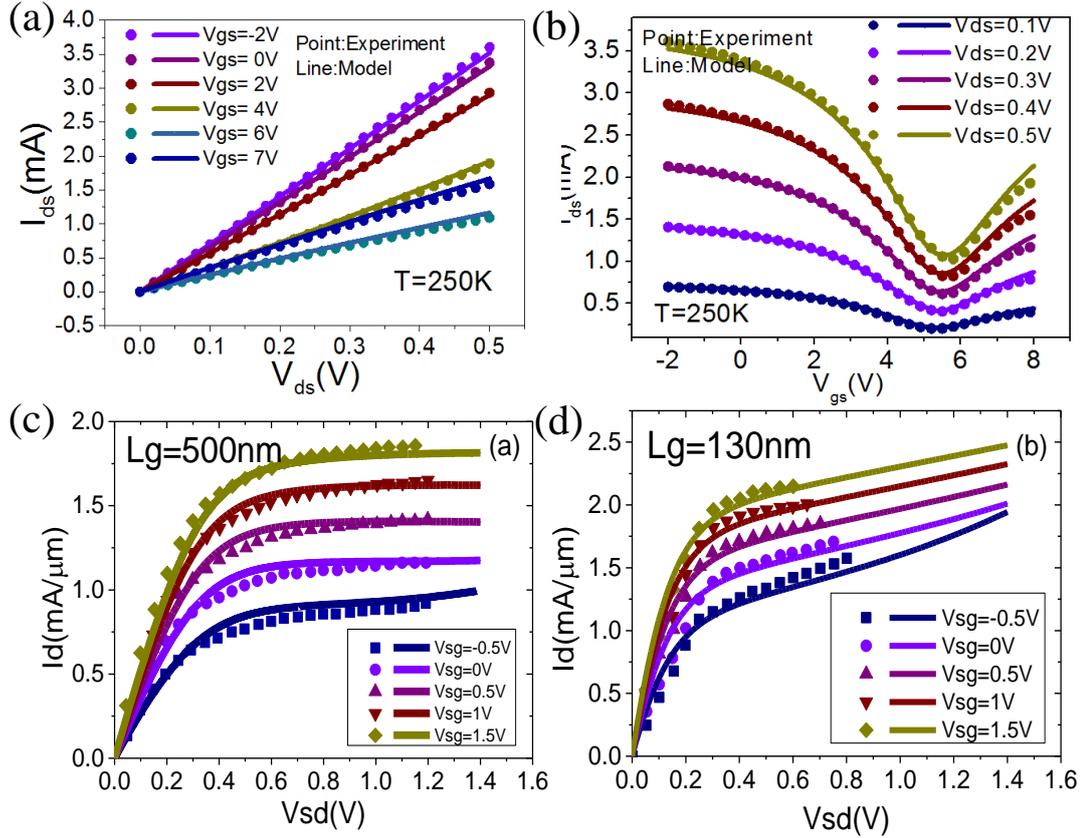

Fig. 15 (a) The output current under different gate voltages for the calculated and experimental results, and (b) transfer current under different drain voltages for the calculated and experimental results. Output current with simulated and experimental resutls under (c) $L_g$=500 nm, and (d) $L_g$=130 nm.

## 7.4 Capacitance model

The terminal charges $Q_g$, $Q_d$ and $Q_s$ always relate with the gate, drain and source electrodes. $Q_g$ is the integration of the net mobile carrier concentration along the channel. $Q_d$ and $Q_s$ can be obtained by Ward-Dutton's linear charge partition scheme [41, 47], with the charge conservation relationship. The expressions for terminal charges are written as

$$\begin{cases} Q_g = -w \int_0^L Q_{net} dy \\ Q_d = w \int_0^L Q_{net} dy \\ Q_s = -(Q_g + Q_d) \end{cases} \cdot \tag{65}$$

Due to the current continuity equation, $dy$ in Eq. (65) can be replaced by $-\mu W |Q_{net}| dV / I_d$ and $y$ is equal to $L_{eff} \frac{I_d(\varphi_s)}{I_d}$. Generally, if one considers the simple current model with constant mobility, disorder induced carrier density can be introduced into Eq. (63). Then, the capacitance model can be deduced analytically by using the formula as

$$C_{ij} = \begin{cases} -\dfrac{\partial Q_i}{\partial v_j} & i \neq j \\ \dfrac{\partial Q_i}{\partial v_j} & i = j \end{cases}, \qquad (66)$$

where $i$ and $j$ represent the terminals. As the gate oxide thickness is greatly decreased, the quantum capacitance needs to account for accurate intrinsic capacitance modeling during circuit simulation. Quantum capacitance induced by disorder is significant in discussing scaling down effect. The comparison of capacitance for the calculated and experimental results is shown in Fig. 16.

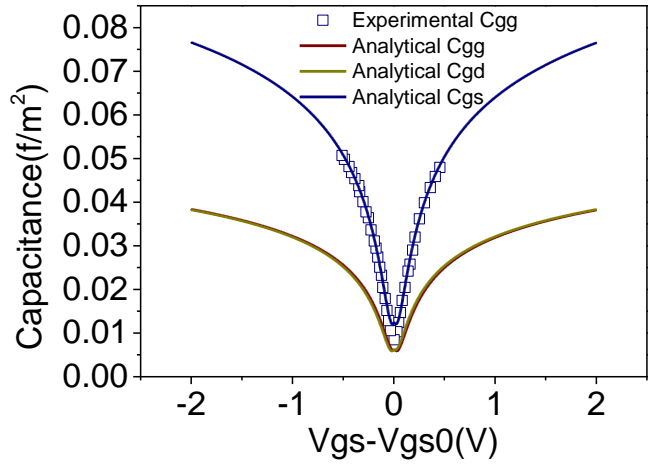

Fig. 16 The comparison of $C_{gg}$, $C_{gs}$ and $C_{gd}$. The lines represent calculated results, and the points represent the experimental results.

## 7.5 Parameter extraction

The computable effective and physics-based extraction method will benefit the compact model's accuracy. Generally, extraction aims at being physical and fitting parameters. To obtain higher level, the parameter sequence should introduce physical effect. However, considering the continuity and accuracy of compact model, the fitting parameters will be used for smoothing the output curves and reducing the error. It is anticipated that the compact model of GFETs with the parameter sets will be suitable to circuit design and can provide accurate insight into the performance. The main criterion for a good set of parameters is the balance of error, efficiency and continuity. Fig. 17 shows the extraction flow of the key physical parameters of the surface-potential-based compact model [47, 48]. Based on the corresponding equations shown in Fig. 17, four key parameters which determine the transport mechanisms can be obtained. Generally speaking, the self-consistently physics-based model is easier to extract the physical parameters. But, it is incompatible to propose a unified method of parameter extraction, due to quite complicated models.

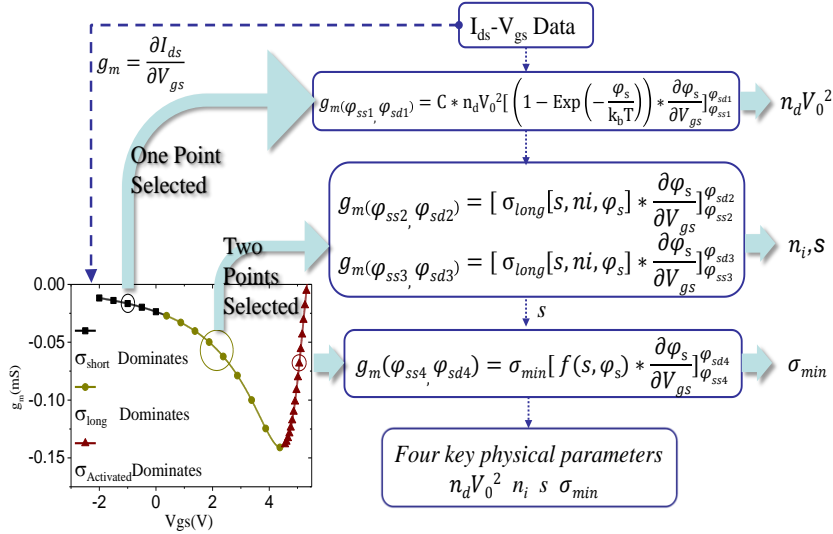

Fig. 17 Extraction flow of the key physical parameters of the model.

## 7.6 Gummel Symmetry Test

For convergence in simulation and analysis for GFET based circuit, continuity and symmetry characteristics have to been kept. On the other hand, to simulate RF circuit, the corresponding parameters should satisfy the continuous and symmetry conditions. Compact model must fulfill one of the benchmark test, that is, Gummel Symmetry Test (GST) [78, 79]. The traditional GST circuit is shown as in Fig. 18. Generally, the higher-order derivatives in MOSFET compact models are obtained as a function of $V_x$, which is symmetry for $V_x$=0. This symmetry roots in the symmetry device structure and channel.

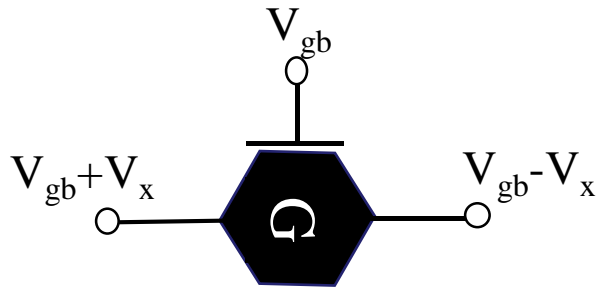

Fig. 18 Traditional GST circuit.

For GFETs, the symmetry is different from the transitional devices, due to the drift of Dirac point. Depletion layer in the channel has disappeared in GFETs. In the capacitance divider scheme, this doping effect causes the asymmetry for plus and minus branches of drain voltage. Thus, to achieve GST, an applied gate-voltage will supplement the drift voltage. Otherwise, to control the continuity, the boundary conditions of the transition of different transport mechanisms cannot be ignored. To achieve it, the current components cover the whole region to avoid the sharp change at some typical points. For $V_x$=0, the hole and electron branches will meet, where smoothing parameter in sgn function will be used for better continuity. If the model passes the GST, as shown in Fig. 19, one can claim that the model has a good continuity and symmetry.

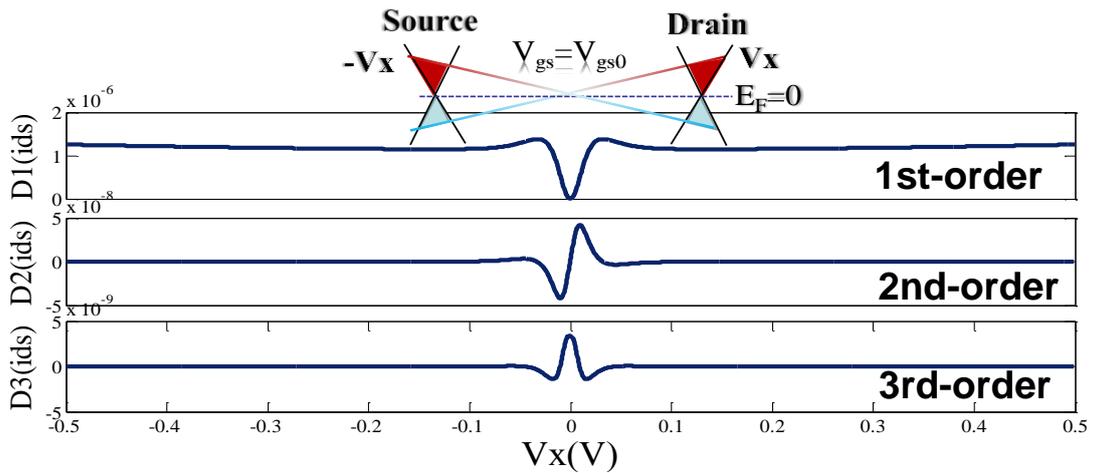

Fig. 19 GST for 1, 2, 3- order derivative of $I_{ds}$ under the condition $V_{gs}=V_{gs0}$.

## 7.7 Simulation of GFET-based circuits

To embed the compact model into a vendor simulator, the Verilog-A (VA) Hardware Description Language (HDL) language is always used for constructing a circuit level model. VA HDL is used for designing analog and mixed-signal systems and integrated circuits defined modules which encapsulate high-level behavioral descriptions as well structural descriptions. Here the surface-potential-based compact model is coded by using Verilog-A language [80]. The parameter declaration, surface-potential calculation, output current, and capacitance are shown in Fig. 20.

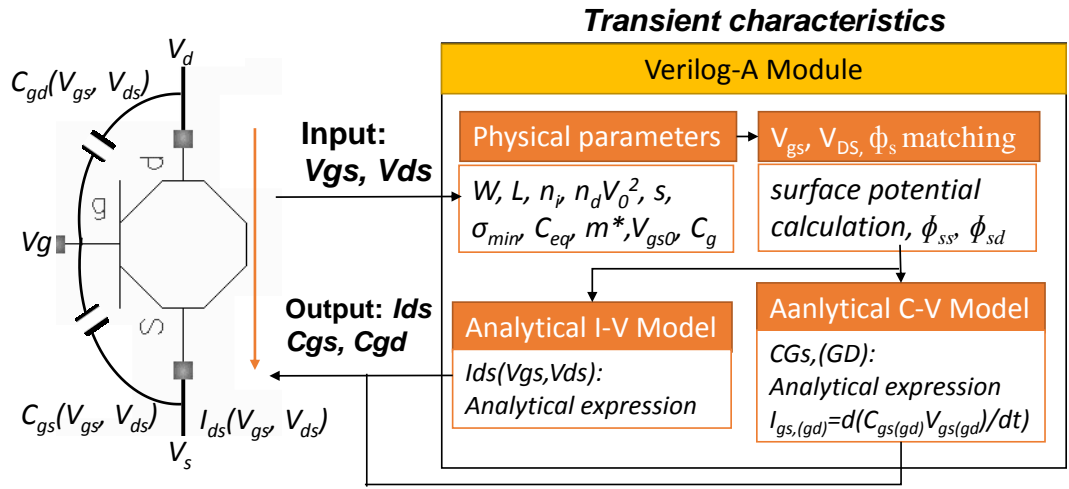

Fig. 20 Verilog-A module for compact model of GFET.

## 8 Conclusions and outlook

In this work, we have reviewed the concept, origin, development and application of compact model within the scope of graphene field-effect transistors (GFETs). Although the compact model has a short history since it appeared, it seems that research interest in the compact model has been growing greatly. Based on the several current contributions, we have tried to describe plenty of methods on developing the GFET compact model. The merits and demerits for current compact model have also been discussed. Otherwise, to keep pace with the increase of circuit operating frequencies and device tolerances scale down with concomitant increases in chip device count, accurate and physics-based compact models are essential for the design and development of GFETs. Currently, the compact model is still open and evolvable. We hope that this review will be helpful to comprehensive understanding of GFETs compact model, and provide a motivation for promoting the application of GFETs in industry.

## Acknowledgment

This work was supported in part by the Opening Project of Key Laboratory of Microelectronics Devices and Integrated Technology, Institute of Microelectronics, Chinese Academy of Sciences, in part by the National Natural Science Foundation of China under Grant 61574166, in part National 973 Program under Grant 2013CBA01604, and in part by National key research and development program (Grant No. 2016YFA0201802), by the Beijing Training Project for the Leading

Talents in S&T under Grant No. Z151100000315008.